\documentclass{pasj00}
\draft

\begin{document}
\SetRunningHead{Fujita et al.}{The high metallicity of ICM}
\Received{2007/03/25}%{yyyy/mm/dd}
\Accepted{2007/05/12}%{yyyy/mm/dd}

\title{High Metallicity of the X-Ray Gas up to the Virial Radius of a
Binary Cluster of Galaxies: Evidence of Galactic Superwinds at
High-Redshift}

%%% begin:list of authors
% Do NOT capitalize all letters in "textsc".
\author{Yutaka \textsc{Fujita}\altaffilmark{1},
Noriaki \textsc{Tawa}\altaffilmark{1},
Kiyoshi \textsc{Hayashida}\altaffilmark{1},
Motokazu \textsc{Takizawa}\altaffilmark{2},\\
Hironori \textsc{Matsumoto}\altaffilmark{3},
Nobuhiro \textsc{Okabe}\altaffilmark{4}, and
Thomas.~H. \textsc{Reiprich}\altaffilmark{5}
}
\altaffiltext{1}{Department of Earth and Space Science, Graduate School of
Science, \\Osaka University, Toyonaka, Osaka 560-0043}
\email{fujita@vega.ess.sci.osaka-u.ac.jp}
\altaffiltext{2}{Department of Physics, Yamagata University, Yamagata
990-8560}
\altaffiltext{3}{Department of Physics, Kyoto University, Kitashirakawa,
Sakyo-ku, Kyoto 606-8502}
\altaffiltext{4}{Astronomical Institute, Graduate School of Science,
Tohoku University, Sendai 980-8578}
\altaffiltext{5}{Argelander Institute for Astronomy (AIfA), Bonn
University, \\ Auf dem H\"ugel 71, 53121 Bonn, Germany}

%% `\KeyWords{}' always has to be placed before `\maketitle'.

\maketitle

\begin{abstract}
 We present an analysis of a Suzaku observation of the link region
 between the galaxy clusters A~399 and A~401. We obtained the
 metallicity of the intracluster medium (ICM) up to the cluster virial
 radii for the first time. We determine the metallicity where the virial
 radii of the two clusters cross each other ($\sim 2$~Mpc away from
 their centers) and found that it is comparable to that in their inner
 regions ($\sim 0.2\: Z_\odot$). It is unlikely that the uniformity of
 metallicity up to the virial radii is due to mixing caused by a cluster
 collision. Since the ram-pressure is too small to strip the
 interstellar medium of galaxies around the virial radius of a cluster,
 the fairly high metallicity that we found there indicates that the
 metals in the ICM are not transported from member galaxies by
 ram-pressure stripping. Instead, the uniformity suggests that the
 proto-cluster region was extensively polluted with metals by extremely
 powerful outflows (superwinds) from galaxies before the clusters
 formed. We also searched for the oxygen emission from the warm--hot
 intergalactic medium in that region and obtained a strict upper
 limit of the hydrogen density ($n_{\rm H}<4.1\times 10^{-5}\rm\:
 cm^{-3}$).
\end{abstract}

\section{Introduction}

Clusters of galaxies are filled with hot X-ray gas ($\sim 2$--10~keV),
which is often called the intracluster medium (ICM) \citep{sar86}. In
the inner region of clusters ($r\lesssim 0.4\: r_{\rm vir}$, where
$r_{\rm vir}$ is the virial radius), the average metallicity of the ICM
is $\sim 0.3\: Z_\odot$ \citep{arn92,fuk00}. The metals in the ICM were
originally produced by stars in galaxies. However, it is still unclear
how and when they were transported into the hot gas from the
galaxies. Possible mechanisms that transfer metals from the cluster
galaxies into the surrounding gas can be classified broadly into two
types, ram-pressure stripping and energetic outflows from the
galaxies. In the former, a galaxy is moving in the ICM and the
metal-enriched gas in the galaxy (interstellar medium) is stripped by
ram-pressure from the ICM \citep{gun72,fuj99,qui00}. The larger is the
density of the ambient ICM and/or the relative velocity between the
galaxy and the ICM, the larger is the ram-pressure affecting the
galaxy. Thus, ram-pressure stripping is most effective at the cluster
center. In the latter, supernova explosions following active star
formation in a galaxy drive outflows, and the metals are carried by them
\citep{dey78}. Since the static pressure from the ICM suppresses the
development of the outflows, the latter mechanism is rather effective in
the peripheral region of a cluster, or the intergalactic space before
the cluster forms (a proto-cluster region) \citep{kap06}. Therefore, it
is critical to determine the metallicity of the ICM in the outermost
region ($r\sim r_{\rm vir}$) of clusters in order to know which
mechanism is dominant. Unfortunately, previous observations of
metallicity have been limited to the inner region of clusters
($r\lesssim 0.4\: r_{\rm vir}$) \citep{deg04,pra07}.

In order to study the nature of the ICM far away from cluster centers,
we observed the link region between two clusters, A~399 and A~401. Their
redshifts are $z=0.0718$ (A~399) and 0.0737 (A~401) \citep{oeg01}. In
this paper, we assume cosmological parameters of $\Omega_0=0.3$,
$\lambda=0.7$, and $H_0=70\rm\: km\: s^{-1}\: Mpc^{-1}$. For those
parameters, the projected distance between the two clusters is 3~Mpc.
Previous X-ray observations showed that these clusters are massive, and
the temperatures are 7.23~keV (A~399) and 8.47~keV (A~401)
\citep{sak04}. They indicated that the clusters are at an initial stage
of a cluster merger, and that the hot gas in the link region is slightly
compressed \citep{sak04,fuj96}.  Since Suzaku has a large collecting
area and low background \citep{mit07}, it is the best instrument for
observations of dim and diffuse X-ray emission, such as that from the ICM
in the periphery of a cluster.

It is believed that there is a warm--hot intergalactic medium (WHIM)
around clusters, which is theoretically predicted, and is known as a
candidate for the ``missing baryon'' in the Universe
\citep{cen99}. While there have been several reports claiming 
detection \citep{fin03,kaa03,fuj04,nic05,tak07b}, there have been
arguments against them \citep{bre06,kaa06,tak07a}.  One difficulty in
detecting the emission or absorption of the WHIM is discrimination
from the emission or absorption of the Galactic warm gas. Since Suzaku
has a good energy resolution of $\Delta z\sim 0.01$ even in the energy
band of $<1$~keV, it can in principal discriminate extragalactic WHIM at
$z\sim 0.07$ from the Galactic warm gas at $z=0$ in redshift space.

This paper is organized as follows. The observations are presented in
section~2. In section~3, the metallicity of the ICM in the outermost
region of the clusters and the line emission from the WHIM are
considered. Section~4 is devoted to discussion. Conclusions are
summarized in section~5.

\section{Observations}

\begin{figure}
  \begin{center}
    \FigureFile(80mm,80mm){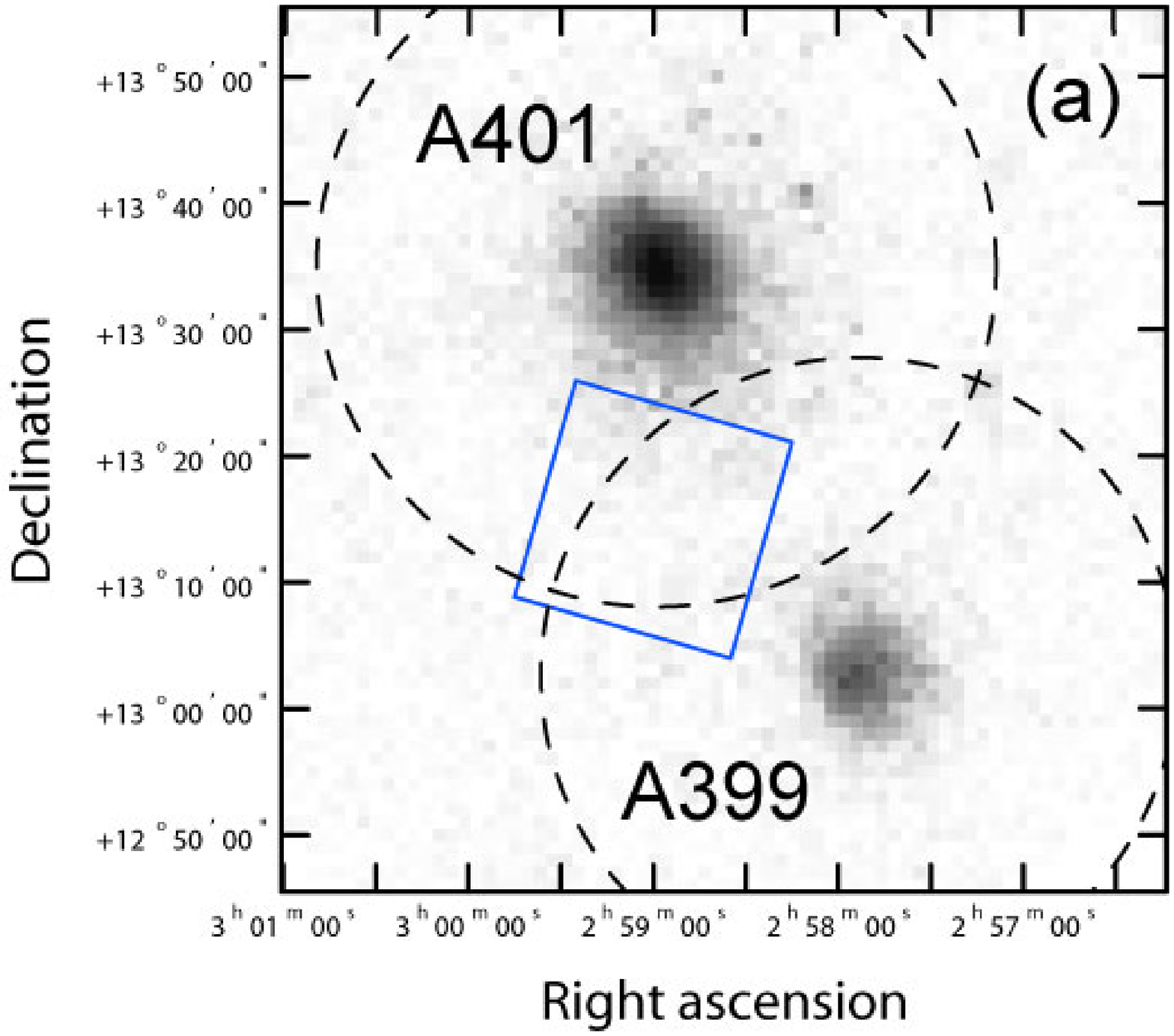}\FigureFile(80mm,80mm){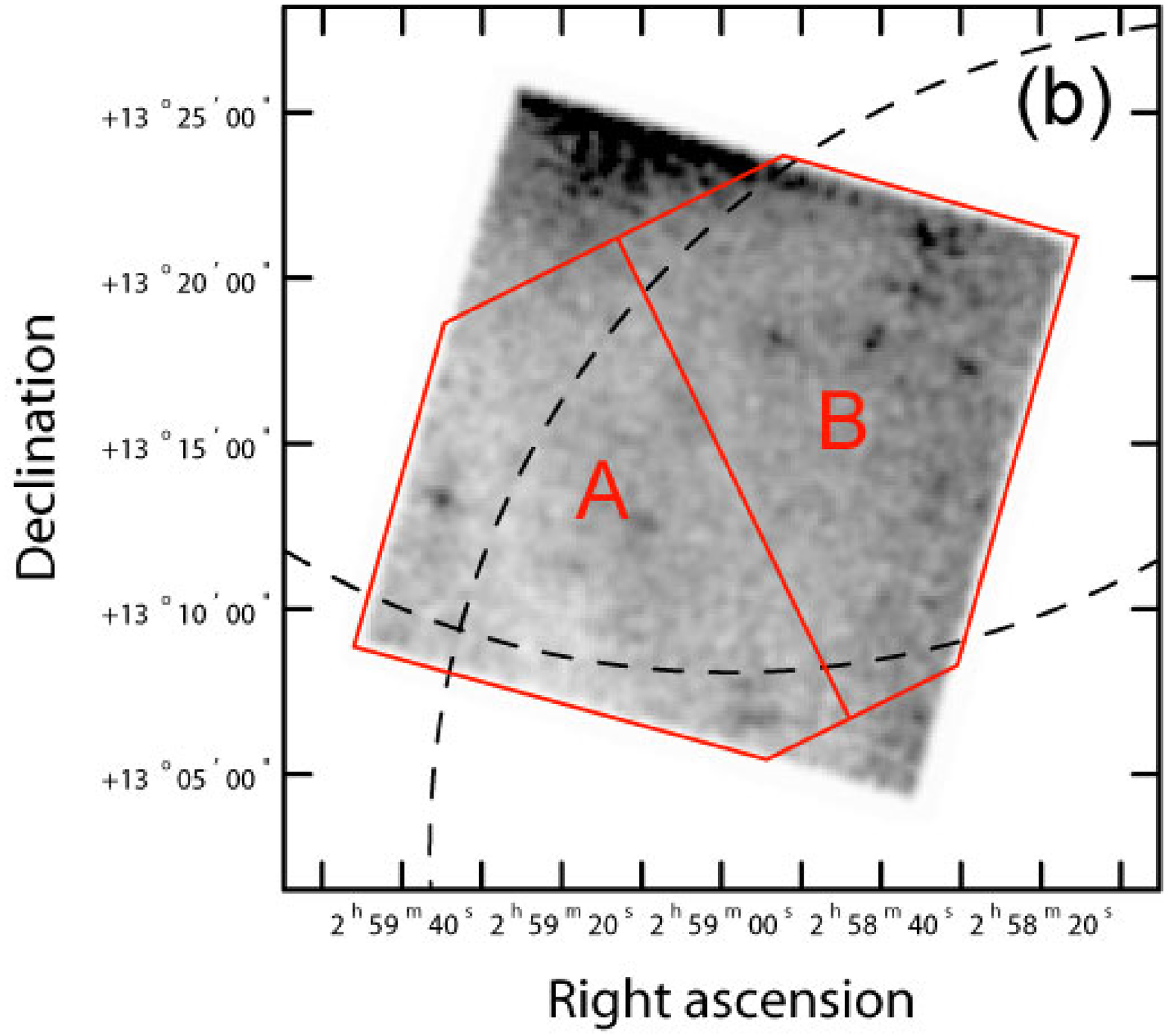}
  \end{center}
  \caption{(a) ROSAT PSPC image of A~399/A~401. The virial radii of the
clusters are shown by dashed-line circles. The Suzaku field shown in
figure~\ref{fig:image}b is indicated by a solid-line square. (b) Suzaku
image of the link region. The background and vignetting are
corrected. The virial radii of the clusters are shown by dashed-line
circles. Regions used for spectral analysis are labeled ``A'' and
``B''.}\label{fig:image}
\end{figure}

We observed the link region with Suzaku on 2006 August 19--22 for an
exposure time of 150~ks. The field of view of Suzaku XIS CCDs is shown
in a ROSAT PSPC image (figure~\ref{fig:image}a). The region that we
observed is the outermost region of the two clusters, where their virial
radii cross each other. The virial radii are approximated by those
inside which the mean mass density is 200-times the critical density of
the Universe. They are $r_{\rm vir}=2.16$~Mpc for A~399 and 2.34~Mpc for
A~401 \citep{sak04}. Because of the projection and the interaction of
the two clusters, the region is brighter than the other sides of the
clusters for a given a distance from the cluster centers
\citep{fuj96,sak04}. The gas that was in the cosmological filament that
connected the two clusters might have been compressed.

Suzaku has four XIS CCDs \citep{koy07}. Three of them are
front-illuminated (FI) and one is back-illuminated (BI). The former have
a higher sensitivity at the high -nergy band ($\gtrsim 2$~keV) than the
latter, and the latter has a higher sensitivity at the low energy band
($\lesssim 2$~keV). Therefore, we used the three FIs to study the hot
ICM (subsection~\ref{sec:ICM}) and the BI to study the WHIM
(subsection~\ref{sec:whim}).

The XIS was operated in the normal full-frame clocking mode.  The edit
mode was $3\times 3$ and $5\times 5$, and we combined the data of both
modes for our analysis. We employed revision 1.2 of the cleaned event
data. Events with ASCA grades of 0, 2, 3, 4, and 6 were retained.  We
excluded the data obtained at the South Atlantic Anomaly, during Earth
occultation, and at the low elevation angles from Earth rims of
$<5^\circ$ (night Earth) and $<20^\circ$ (day Earth).
Figure~\ref{fig:image}b is the summed image of the three FIs
(0.4--5~keV).  Although the north (the region closer to A~401) is
brighter than the south-east corner, noticeable structures, such as a
group of galaxies, are not seen.

\section{Spectral Analysis}
\label{sec:spec}

\subsection{The ICM in the Link Region}
\label{sec:ICM}

We extracted spectra for regions A and B (see figure~\ref{fig:image}b)
in the 0.4--10~keV band. Since the X-rays from the link region is
diffuse and extended, we calculated the effective area for each XIS chip
using XISSIMARFGEN (version 2006-11-26), which provides the ancillary
response file (ARF) through Monte Carlo simulations. Based on the beta
model parameters and temperatures \citep{sak04}, we also calculated the
contamination of photons from the bright centers of the clusters by
simulating the number of photons from outside of the Suzaku
field. Although we found that it is only 4\% of the total photons in the
Suzaku field, we included the effect of the contaminating photons in the
ARF.

Non-X-ray instrumental background (NXB) was subtracted from the spectrum
of the link region using night Earth data. However, the NXB data
provided by the Suzaku team were based on the night Earth data taken
from September 2005 to May 2006, and we found that some corrections were
required to apply them to our observations made in August 2006
\citep{taw08}. The problem is caused by the attenuation of metal lines
scattered from the calibration source and the degradation of spectral
resolution of XIS.

First, we fit the lines in the NXB spectrum with Gaussian profiles using
the redistribution matrix file (RMF) of 2005 August, which was suitable
for the spectral analysis of data taken in the period when the night
Earth data were taken. The normalization and width of the lines were not
fixed in the fit. Next, we subtracted the lines from the NXB spectrum
and obtained the continuum spectrum of the NXB. For the Mn lines from
the calibration source, we considered the radioactive decay of Mn (the
half period is 2.73~yr). For prominent lines, we consider the
degradation of the spectral resolution of XIS. Then, we simulated the
line profiles in 2006 August. Finally, adding the simulated lines to the
continuous spectrum of the NXB, we obtain the corrected NXB background;
we used it for the following spectral analysis.

\begin{figure}
  \begin{center}
    \FigureFile(120mm,120mm){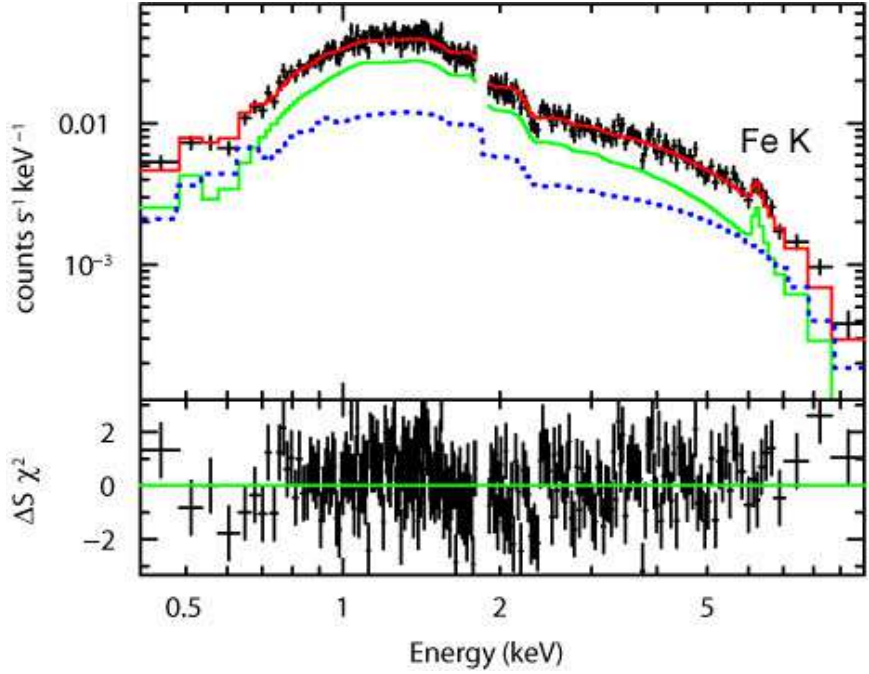}
  \end{center}
  \caption{X-ray spectrum of region A shown in figure~\ref{fig:image}b
(crosses). Three FI XIS spectra were summed. The result of the fit is
shown by the red line, while the lower panel plots the residuals divided
by the $1\sigma$ errors. The green solid line shows the contribution of
the ICM, and the blue dotted line shows the contribution of CXB and
Galactic emission. The excess seen at $\gtrsim 6$~keV, which is also
seen in figure~\ref{fig:spec_whim}, may be due to the field-to-field
variation of the CXB strength (see text).}\label{fig:spec}
\end{figure}

In the spectral analysis, X-ray point sources in the field were
removed. For each region (A and B), the spectra of the three FIs were
summed. Using XSPEC (version 12), the summed spectra were fitted with a
single thermal model (APEC) representing the ICM and with Galactic
absorption (WABS). While the redshift of the ICM ($z=0.073$) and the
Galactic absorption ($N_{\rm H}=1\times 10^{21}\rm\: cm^{-2}$;
\cite{dic90}) were fixed in the fits, the temperature and metallicity of
the ICM were free. In the fits, we also considered the contributions of
the cosmic X-ray background (CXB) obtained by ASCA \citep{kus02} and the
Galactic soft X-ray
emission\footnote{http://heasarc.gsfc.nasa.gov/cgi-bin/Tools/xraybg/xraybg.pl.}.
The CXB spectrum is given by a power-law with an index of 1.412, and the
flux in the 2--10~keV band is $F_{\rm CXB}=6.38\times 10^{-8}\rm\: erg\:
cm^{-2}\: s^{-1}\: sr^{-1}$. We call this CXB spectrum `the fiducial CXB
spectrum'. The Galactic soft X-ray emission consists of two thermal
components of $T=0.0827$ (the local hot bubble) and 0.184~keV (the Milky
Way halo) \citep{sno98}. The metallicity and the redshift of the both
components are one solar abundance and zero, respectively. The
parameters for the Galactic soft X-ray emission were fixed in the fits,
except for the overall normalization; the relative normalization between
the two components was fixed at 1.04, which is provided by the Galactic
soft X-ray emission model$\:^1$. The following results about the hot ICM
are not sensitive to the Galactic soft X-ray
emission. Figure~\ref{fig:spec} shows the summed spectrum of the three
FIs for region A and the result of the fit; an Fe~K line is clearly
seen. We fixed the CXB spectrum at the fiducial one. For this fit,
$\chi^2/{\rm dof}=1844.60/1731$.

\begin{figure}
  \begin{center}
    \FigureFile(120mm,120mm){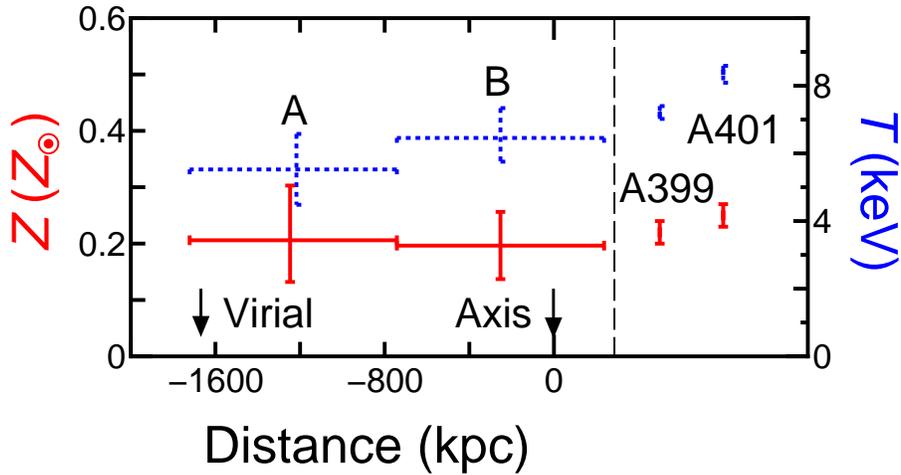}
    %%% \FigureFile(width,height){filename}
  \end{center}
  \caption{Temperatures (blue, dotted) and metallicities (red, solid) of
regions A and B. Those of the inner regions of A~399 and A~401 are also
shown for comparison \citep{sak04}. The abscissa axis is the distance
from the line connecting the two cluster centers (``Axis''). The
position where the virial radii of the two clusters cross each other is
referred as ``Virial''. The error bars are 90\% confidence
intervals.}\label{fig:TZ}
\end{figure}

In figure~\ref{fig:TZ}, we present the temperatures and metallicities of
regions A and B.  When we derived the error bars, we varied the
normalization of the CXB in order to involve any possible field-to-field
variance. \citet{kus02} estimated that the intensity changed
$6.49_{-0.61}^{+0.56}$\% for the ASCA field ($\sim 0.5\rm\: deg^2$). The
area of regions A and B is $\sim 0.04\rm\: deg^2$. Assuming that the CXB
intensity follows the Poisson model, the CXB uncertainty for regions A
and B is $\sim 6.5\sqrt{0.5/0.04}\approx 23$\%. Conservatively, we take
an uncertainty of 30\%. We fit the XIS spectra of regions A and B for
30\% higher or lower CXB normalizations than that of the fiducial CXB
spectrum and derive temperatures and metallicities. Adding the
statistical errors of the fits to this CXB uncertainty, we obtained the
error bars in figure~\ref{fig:TZ}.

In that figure, we compare the temperatures and metallicities in the link
region with the average ones of the inner regions of A~399 and A~401
($r\lesssim 0.4\: r_{\rm vir}$) \citep{sak04}. The temperatures of
regions A and B would have been raised slightly from their original
values because of gas compression associated with the interaction
between the two clusters. The metallicities of regions A and B ($\sim
0.2\: Z_\odot$) are not different from those in their inner regions.

\subsection{Warm-Hot Intergalactic Medium}
\label{sec:whim}

The high ICM metallicity observed in the link region ($\sim 0.2\:
Z_\odot$) may indicate that the WHIM around clusters would also be
highly polluted with metals \citep{cen06}. Thus, the metal line emission
from the WHIM could be detected \citep{yos03}. We searched for
O\emissiontype{VII} line emission from the WHIM around the binary
cluster superposed on the Suzaku field.

At the redshift of the link region ($z\sim 0.073$), the
O\emissiontype{VII} line (intrinsically 0.57~keV) would be observed at
0.53~keV. Unfortunately, X-ray spectra in the energy band of $\lesssim
1$~keV are often affected by the solar wind \citep{fuj07}. The level 2
ACE SWEPAM
data\footnote{http://swepam.lanl.gov/data/raw/index.cgi/swepam\_dswi\_level2.}
show that in the first half of our observation time, the proton flux is
$\sim 5\times 10^8\rm\: cm^{-2}\: s^{-1}$, which is comparable to the
level when solar-wind charge-exchange X-ray emission from the Earth's
magnetosheath was observed \citep{fuj07}. In fact, we found that the XIS
photon counts in the 0.3--1~keV band are $\sim 20$\% larger in the first
half of our observation time than those in the second half. Thus, we
used only the data of the second half of our observation time (75~ks),
because they are less affected by the solar wind. We apply the analysis
in subsection~\ref{sec:ICM} to the BI.

The BI spectrum of the entire Suzaku field (figure~\ref{fig:image}) was
fitted in the 0.25--8~keV band with a single thermal model (APEC), a
Gaussian corresponding to the expected O\emissiontype{VII} line around
$0.53$~keV, and other components not relating to the binary cluster (the
CXB, the Galactic soft X-ray emission, and the Galactic absorption). The
thermal component is required to precisely estimate the contribution of
the hot ICM. We assumed that the metallicity and redshift of the thermal
component are $Z=0.2\: Z_\odot$ and $z=0.073$, respectively. The former
is the ICM metallicity derived with FI (subsection~\ref{sec:ICM}). The
temperature and normalization of the thermal component were free in the
fit. While the unknown intrinsic width of the Gaussian component was
fixed at zero, the line center and normalization of the component were
not fixed.  The parameters for the CXB, the Galactic soft X-ray
emission, and the Galactic absorption were the same as those in the
previous analysis for the FIs. The overall normalization of the Galactic
soft X-ray emission was varied. When we derived error bars for the
temperature of the thermal component and the strength of the
O\emissiontype{VII} line emission from the WHIM, the 30\% uncertainty of
the CXB was considered.

\begin{figure}
  \begin{center}
    \FigureFile(120mm,120mm){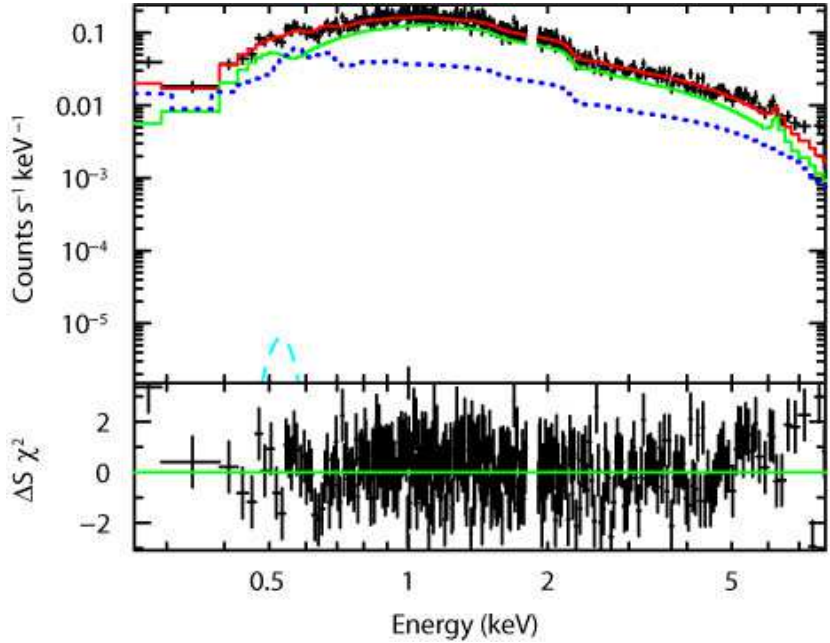}
  \end{center}
  \caption{Same as figure~\ref{fig:spec} but for the BI spectrum of the
 entire Suzaku field. The contribution of the Gaussian component shown
 at 0.53~keV is that when the line center energy was fixed at
 $0.53$~keV. }\label{fig:spec_whim}
\end{figure}

The result of the fit is shown in figure~\ref{fig:spec_whim}. For this
fit, $\chi^2/{\rm dof}=524.98/477$. The temperature of the thermal
component is $T=5.8^{+0.9}_{-0.9}$~keV. No line is seen at
0.53~keV. Another thermal component that may represent the WHIM is not
required. Although we have tried to fit the FI and BI spectra
simultaneously, the results were not improved.

\section{Discussion}
\label{sec:dis}

\subsection{The High Metallicity of the ICM}

Figure~\ref{fig:TZ} suggests that the metallicity of the ICM is uniform
up to the virial radii of the two clusters.  Since the surface
brightness of a cluster is a decreasing function of the radius, the
fraction of photons coming from the virial radii may not be large in the
Suzaku field.  However, they come from at least $r\geq 0.5\: r_{\rm
vir}$ (see Fig.~\ref{fig:image}), where the metallicity has not been
obtained before.  The uniformity could be due to a head-on collision of
the two clusters; if the two clusters had already passed through each
other, metal-rich gas would have been pulled out of their central
regions. However, numerical simulations showed that while collisionless
dark matter and galaxies can pass through the other cluster at a cluster
collision, collisional ICM cannot. Thus, the ICM is detached from the
dark matter and galaxies \citep{tak99,ric01,poo06}. However, X-ray
observations have shown that the overall X-ray morphology of the two
clusters is regular (figure~\ref{fig:image}a), which indicates that the
ICM is almost in pressure equilibrium in the gravitational potentials
formed by the dark matter \citep{sak04}. Moreover, their X-ray centers
are coincident with their central galaxies as is the case of non-merging
clusters \citep{fuj96}. These indicate that the clusters have not passed
through each other. Although both A~399 and A~401 do not have a
prominent cool core, Sakelliou and Ponman (2004) concluded that this is
because each cluster has undergone minor mergers, which did not
significantly affect the overall structures. Still, we cannot rule out
that the minor mergers might have resulted in some mixing of metals. It
is, therefore, important that the findings here also be confirmed with
classical relaxed clusters. Moreover, we determined only the metallicity
of the link region. Those on the other sides of both clusters should be
determined in the future.

It is also unlikely that the high metallicity observed with Suzaku is
due to a small-scale structure with high
metallicity. Figure~\ref{fig:image} shows that there is no noticeable
X-ray structure in the Suzaku field. Moreover, there is no enhancement
of the number density of galaxies in that region \citep{sak04}. These
results mean that there is no group of galaxies that possibly ejects
more metals than the surrounding region. Furthermore, there is no
difference of metallicity between regions A and B
(figure~\ref{fig:TZ}). This shows that the metallicity is uniform at
least on a scale of $\sim 1$~Mpc.

If the metallicity is actually high in the outermost region other than
the link region, and if the high metallicity is not due to cluster
mergers, it constrains models of metal transportation from
galaxies. First, ram-pressure stripping is rejected as the main
mechanism, because the ram-pressure is very small and the stripping is
ineffective at $r\sim r_{\rm vir}$ \citep{fuj99,dom06}. Note that
although the two clusters are interacting, they are just at an initial
stage of a collision.  Using the beta model parameters for both clusters
obtained by Sakelliou and Ponman (2004), the summed ICM density in the
Suzaku field is $\sim 2.4\times 10^{-4}\rm\: cm^{-3}$. Since the surface
brightness in the field is a factor of two larger than the simple
superposition of the two clusters (figure~10 in \cite{sak04}), the
actual density would be $\sqrt{2}$ times larger, and be $\sim 3.4\times
10^{-4}\rm\: cm^{-3}$.  For a typical galaxy, ram-pressure stripping
happens when the ram-pressure exceeds $\sim 2\times 10^{-11}\rm\: dyne\:
cm^{-2}$ \citep{fuj99}.  Thus, the relative velocity between a galaxy
and the ICM must be larger than $\sim 2000\rm\: km\: s^{-1}$ for
effective ram-pressure stripping. It is unlikely that this happens at
this stage of a cluster collision \citep{fuj99b}.

Therefore, outflows from galaxies are the most promising candidates of
the metal transportation mechanism. However, theoretical studies have
shown that simple galactic-outflow models using standard parameters for
star formation do not reproduce the level of uniformity that we
observed. They predict that metallicity in the outermost region of a
cluster should be at least a factor of a few smaller than the average
metallicity in the inner region ($r\lesssim 0.4\: r_{\rm vir}$)
\citep{tor04,kap06}. In particular, metal maps obtained by the latest
numerical simulations [e.g. figures~5 and 6 of \citet{kap07}] showed
that metals are concentrated in the central regions ($\lesssim 1$~Mpc)
of clusters with temperatures of $T\sim 8$~keV, except for small
substructures with a size of $\sim 0.5$~Mpc. The reason is that in these
models most metals are produced by galaxies within an already formed
cluster potential well and the ICM at $z<1$--2.  Thus, it is difficult
for ordinary outflows to transfer metals away from galaxies. In
particular, elliptical galaxies, which are often thought to be the main
metal sources of the ICM \citep{arn92}, are concentrated in the central
regions of clusters at $z\sim 0$ (e.g. \cite{got03}). It is not feasible
for the elliptical galaxies in the grown-up clusters to pollute the ICM
at $r\sim r_{\rm vir}$ with metals. For example, an elliptical galaxy
with a luminosity of $L_{\rm B}=10^{10}\: L_\odot$ releases an energy of
$E_{\rm w}\sim 10^{60}$~erg through galactic outflow \citep{dav91}.  The
distance from the galaxy to which the outflow can reach, $d_{\rm w}$,
has the relation
\begin{equation}
 E_{\rm w}\sim (4\pi/3)P d_{\rm w}^3\:,
\end{equation}
where $P$ is the pressure of the ICM surrounding the galaxy. Thus,
\begin{equation}
 d_{\rm w}\sim  86
\left(\frac{n}{10^{-3}\rm\: cm^{-3}}\right)^{-1/3}
\left(\frac{T}{8\rm\: keV}\right)^{-1/3}
\left(\frac{E_{\rm w}}{10^{60}\rm\: erg}\right)^{1/3}\:{\rm kpc}\:,
\end{equation}
where $n$ is the density of the ICM surrounding the galaxy. The distance
$d_{\rm w}$ is much smaller than the virial radius of a cluster ($\sim
2$~Mpc), which shows that the galactic outflows from elliptical galaxies
at $z\sim 0$ alone cannot explain the high metallicity observed at $r\sim
r_{\rm vir}$.

The solution may be extremely powerful outflows from galaxies at
high-redshifts ($z\sim 2$). At that time, clusters had not much grown,
and most galaxies observed in clusters at $z\sim 0$ had not fallen into
the clusters. Such outflows (superwinds) could be produced by active
galactic nuclei (AGNs) or intensive starburst activities
\citep{ben03,cen06}, and they could pollute gas with metals throughout
the proto-cluster region \citep{rom06,mol07}. That metal-enriched gas
would be captured by A~399 and A~401 recently, and would be observed as
the ICM at present. In fact, the space density of luminous AGNs and that
of starburst galaxies increase from $z=0$ to $z\sim 2$
\citep{ued03,fra01}. Using semi-analytical galaxy formation models,
\citet{nag05} indicated that the metals were ejected into space outside
the cluster ancestors by superwinds before the circular velocity of the
individual ancestors increased to $\sim 600\:\rm km\: s^{-1}$.

\subsection{Warm-Hot Intergalactic Medium}

\begin{figure}
  \begin{center}
    \FigureFile(80mm,80mm){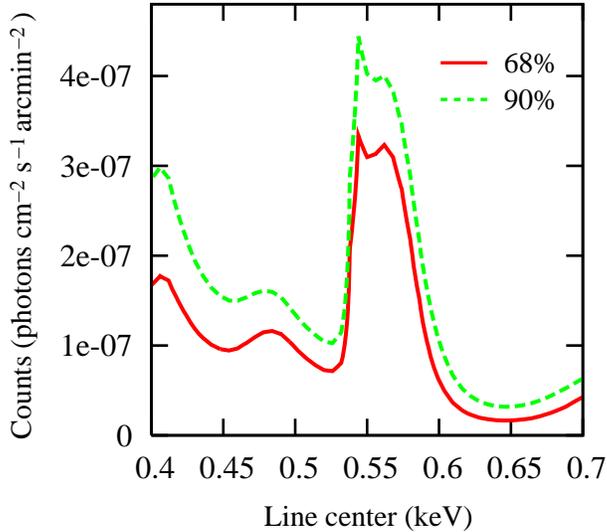}
  \end{center}
  \caption{Confidence contours for the line center and photon counts of
the Gaussian component. The lines show 68\% and 90\% confidence regions
for two interesting parameters. For this figure, the 30\% uncertainty of
the CXB is not considered, because we found that the effect is
negligible.}\label{fig:conf}
\end{figure}

We estimated the upper limit of the O\emissiontype{VII} line emission from
the WHIM. Figure~\ref{fig:conf} shows the confidence contours for the
line center energy and the normalization. While there is no structure
around 0.53~keV, there is a noticeable structure at $\sim
0.57$~keV. Thus, the possible line in the spectrum could be associated
with an insufficient correction of the Galactic soft X-ray emission. If
we assume that the center of the line is at 0.53~keV, the line strength
is $I<8.0\times 10^{-8}\rm\: photons\; cm^{-2}\: s^{-1}\: arcmin^{-2}$
(90\% confidence level). Assuming that the temperature of the gas
associated with the O\emissiontype{VII} line is $T=2\times 10^6$~K, the
hydrogen density at $z=0.073$ can be represented by
\begin{eqnarray}
 n_{\rm H}&=&9.2\times 10^{-5}{\rm cm^{-3}}
\left(\frac{I}{1\times 10^{-7}\rm ph\: cm^{-2}\: s^{-1}}\right)^{1/2}
\left(\frac{Z}{0.1\: Z_\odot}\right)^{-1/2}
\left(\frac{L}{1\rm\: Mpc}\right)^{-1/2}\:,
\label{eq:nH}
\end{eqnarray}
where $L$ is the path length \citep{tak07a}. From
equation~(\ref{eq:nH}), the density of the WHIM is $n_{\rm H}<4.1\times
10^{-5}\rm\: cm^{-3}$ for $Z=0.2\: Z_\odot$ and $L=2$~Mpc. The value of
$Z$ is the ICM metallicity obtained with FIs and that of $L$ is the
typical depth of warm gas in a cosmic filament \citep{col05}.

The allowed density is smaller than that of the hot ICM in the link
region ($n_{\rm H}\sim 3.4\times 10^{-4}\rm\: cm^{-3}$; see
section~\ref{sec:dis}). The upper limit is also smaller than that
obtained for A2218 ($n_{\rm H}<5.5\times 10^{-5}\rm\: cm^{-3}$ for
$Z=0.2\: Z_\odot$, and $L=2$~Mpc; \cite{tak07a}). It is to be noted that
since the emissivity of the O\emissiontype{VII} line is peaked at
$T\approx 2\times 10^6$~K, the hydrogen density could be larger if the
temperature of the WHIM is much different from $T=2\times 10^6$~K.

\section{Conclusions}

We observed the link region between massive clusters A~399 and A~401
with the XIS instrument aboard Suzaku. We measured the metallicity of
the hot ICM where the virial radii of the two clusters cross each other,
and found that it is $\sim 0.2\: Z_\odot$, which is comparable to those
in the inner regions of the clusters. This suggests that the metallicity
is uniform within the virial radii. The X-ray morphology of the two
clusters indicates that the uniformity of the metallicity is not due to
a major cluster merger. The fairly high metallicity in the outermost
region of the clusters shows that ram-pressure stripping is not the main
mechanism that transports metals from galaxies to the hot ICM. Moreover,
the uniformity suggests that the metals were not ejected through
ordinary galactic outflows after the clusters formed. It is more likely
that the metals were carried by galactic superwinds before the clusters
formed, and that the proto-cluster region was heavily polluted with the
metals.

We also searched for the oxygen emission from the WHIM around the
clusters. We could obtain a strict constraint of the density of the
WHIM, which is $n_{\rm H}< 4.1\times 10^{-5}\rm\: cm^{-3}$.

\vspace{5mm}

The authors wish to thank the referee for useful comments.  We also
thank the Suzaku operations team for their support in planning and
executing these observations.  Y.~F., K.~H., and M.~T. were supported in
part by a Grant-in-Aid from the Ministry of Education, Culture, Sports,
Science and Technology of Japan (Y.~F.: 17740162; K.~H.:16002004;
M.~T.: 16740105, 19740096). T.~H.~R. acknowledges support by the
Deutsche Forschungsgemeinschaft through Emmy Noether Research Grant RE
1462.

%%%
% See the manual for the detail.
%%%


\begin{thebibliography}{}

\bibitem[Arnaud et al.(1992)]{arn92} Arnaud, M., Rothenflug, 
R., Boulade, O., Vigroux, L., \& Vangioni-Flam, E.\ 1992, \aap, 254, 49 

\bibitem[Benson et al.(2003)]{ben03} Benson, A.~J., Bower, 
R.~G., Frenk, C.~S., Lacey, C.~G., Baugh, C.~M., \& Cole, S.\ 2003, \apj, 
599, 38 

\bibitem[Bregman, Lloyd-Davies(2006)]{bre06} Bregman, 
J.~N., \& Lloyd-Davies, E.~J.\ 2006, \apj, 644, 167 

\bibitem[Cen, Ostriker(1999)]{cen99} Cen, R., \& Ostriker, 
J.~P.\ 1999, \apj, 514, 1 

\bibitem[Cen, Ostriker(2006)]{cen06} Cen, R., \& Ostriker, 
J.~P.\ 2006, \apj, 650, 560 

\bibitem[Colberg et al.(2005)]{col05} Colberg, J.~M., 
Krughoff, K.~S., \& Connolly, A.~J.\ 2005, \mnras, 359, 272 

\bibitem[David et al.(1991)]{dav91} David, L.~P., Forman, W., 
\& Jones, C.\ 1991, \apj, 380, 39 

\bibitem[De Grandi et al.(2004)]{deg04} De Grandi, S., 
Ettori, S., Longhetti, M., \& Molendi, S.\ 2004, \aap, 419, 7 

\bibitem[De Young(1978)]{dey78} De Young, D.~S.\ 1978, \apj, 
223, 47 

\bibitem[Dickey, Lockman(1990)]{dic90} Dickey, J.~M., \& 
Lockman, F.~J.\ 1990, \araa, 28, 215 

\bibitem[Domainko et al.(2006)]{dom06} Domainko, W., et al.\ 
2006, \aap, 452, 795 

\bibitem[Finoguenov et al.(2003)]{fin03} Finoguenov, A., 
Briel, U.~G., \& Henry, J.~P.\ 2003, \aap, 410, 777 

\bibitem[Franceschini et al.(2001)]{fra01} Franceschini, A., 
Aussel, H., Cesarsky, C.~J., Elbaz, D., \& Fadda, D.\ 2001, \aap, 378, 1 

\bibitem[Fujimoto et al.(2004)]{fuj04} Fujimoto, R., et al.\ 
2004, \pasj, 56, L29 

\bibitem[Fujimoto et al.(2007)]{fuj07} Fujimoto, R., et al.\ 
2007, \pasj, 59, S133

\bibitem[Fujita et al.(1996)]{fuj96} Fujita, Y., Koyama, K., 
Tsuru, T., \& Matsumoto, H.\ 1996, \pasj, 48, 191 

\bibitem[Fujita, Nagashima(1999)]{fuj99} Fujita, Y., \& 
Nagashima, M.\ 1999, \apj, 516, 619 

\bibitem[Fujita et al.(1999)]{fuj99b} Fujita, Y., Takizawa, 
M., Nagashima, M., \& Enoki, M.\ 1999, \pasj, 51, L1 

\bibitem[Fukazawa et al.(2000)]{fuk00} Fukazawa, Y.,
Makishima, K., Tamura, T., Nakazawa, K., Ezawa, H., Ikebe, Y., 
Kikuchi, K., 
\& Ohashi, T.\ 2000, \mnras, 313, 21 

\bibitem[Goto et al.(2003)]{got03} Goto, T., Yamauchi, C., 
Fujita, Y., Okamura, S., Sekiguchi, M., Smail, I., Bernardi, M., \&
		Gomez,
P.~L.\ 2003, \mnras, 346, 601 

\bibitem[Gunn, Gott(1972)]{gun72} Gunn, J.~E., \& Gott, 
J.~R., III\ 1972, \apj, 176, 1 

\bibitem[Kaastra et al.(2003)]{kaa03} Kaastra, J.~S., Lieu, 
R., Tamura, T., Paerels, F.~B.~S., \& den Herder, J.~W.\ 2003, \aap, 397, 
445 

\bibitem[Kaastra et al.(2006)]{kaa06} Kaastra, J.~S., Werner, 
N., den Herder, J.~W.~A., Paerels, F.~B.~S., de Plaa, J., Rasmussen,
		A.~P.,
\& de Vries, C.~P.\ 2006, \apj, 652, 189 

\bibitem[Kapferer et al.(2006)]{kap06} Kapferer, W., et al.\ 
2006, \aap, 447, 827 

\bibitem[Kapferer et al.(2007)]{kap07} Kapferer, W., et al.\ 
2007, \aap, 466, 813 

\bibitem[Koyama et al.(2007)]{koy07} Koyama, K., et al.\ 
2007, \pasj, 59, S23 

\bibitem[Kushino et al.(2002)]{kus02} Kushino, A., Ishisaki, 
Y., Morita, U., Yamasaki, N.~Y., Ishida, M., Ohashi, T., \& Ueda, Y.\ 2002, 
\pasj, 54, 327 

\bibitem[Mitsuda et al.(2007)]{mit07} Mitsuda, K., et al.\ 
2007, \pasj, 59, S1 

\bibitem[Moll et al.(2007)]{mol07} Moll, R., et al.\ 2007, 
\aap, 463, 513 

\bibitem[Nagashima et al.(2005)]{nag05} Nagashima, M., Lacey, 
C.~G., Baugh, C.~M., Frenk, C.~S., \& Cole, S.\ 2005, \mnras, 358, 1247 

\bibitem[Nicastro et al.(2005)]{nic05} Nicastro, F., et al.\ 
2005, \nat, 433, 495

\bibitem[Oegerle, Hill(2001)]{oeg01} Oegerle, W.~R., \& 
Hill, J.~M.\ 2001, \aj, 122, 2858 

\bibitem[Poole et al.(2006)]{poo06} Poole, G.~B., Fardal, 
M.~A., Babul, A., McCarthy, I.~G., Quinn, T., \& Wadsley, J.\ 2006, \mnras, 
373, 881 

\bibitem[Pratt et al.(2007)]{pra07} Pratt, G.~W.,
B{\"o}hringer, H., Croston, J.~H., Arnaud, M., Borgani, S., 
Finoguenov, A., 
\& Temple, R.~F.\ 2007, \aap, 461, 71 

\bibitem[Quilis et al.(2000)]{qui00} Quilis, V., Moore, B., 
\& Bower, R.\ 2000, Science, 288, 1617 

\bibitem[Ricker, Sarazin(2001)]{ric01} Ricker, P.~M., \& 
Sarazin, C.~L.\ 2001, \apj, 561, 621 

\bibitem[Romeo et al.(2006)]{rom06} Romeo, A.~D., 
Sommer-Larsen, J., Portinari, L., \& Antonuccio-Delogu, V.\ 2006, \mnras, 
371, 548 

\bibitem[Sakelliou, Ponman(2004)]{sak04} Sakelliou, I., \& 
Ponman, T.~J.\ 2004, \mnras, 351, 1439 

\bibitem[Sarazin(1986)]{sar86} Sarazin, C.~L.\ 1986, Rev. Mod. Phys.,
		58, 1

\bibitem[Snowden et al.(1998)]{sno98} Snowden, S.~L., Egger, 
R., Finkbeiner, D.~P., Freyberg, M.~J., \& Plucinsky, P.~P.\ 1998, \apj, 
493, 715 

\bibitem[Takei et al.(2007a)]{tak07a} Takei, Y., et al.\ 2007a, 
\pasj, 59, S339 

\bibitem[Takei et al.(2007b)]{tak07b} Takei, Y., Henry, J.~P., 
Finoguenov, A., Mitsuda, K., Tamura, T., Fujimoto, R., \& Briel, U.~G.\ 
2007b, \apj, 655, 831 

\bibitem[Takizawa(1999)]{tak99} Takizawa, M.\ 1999, \apj, 
520, 514 

\bibitem[Tawa et al.(2008)]{taw08} Tawa, N., et al.\ 2008, \pasj, 60,
		S11

\bibitem[Tornatore et al.(2004)]{tor04} Tornatore, L., 
Borgani, S., Matteucci, F., Recchi, S., \& Tozzi, P.\ 2004, \mnras, 349, 
L19 

\bibitem[Ueda et al.(2003)]{ued03} Ueda, Y., Akiyama, M., 
Ohta, K., \& Miyaji, T.\ 2003, \apj, 598, 886 

\bibitem[Yoshikawa et al.(2003)]{yos03} Yoshikawa, K., 
Yamasaki, N.~Y., Suto, Y., Ohashi, T., Mitsuda, K., Tawara, Y., \& 
Furuzawa, A.\ 2003, \pasj, 55, 879 

% Journals(e.g. A\&A,ApJ,AJ,NMRAS,PASP ...)
% Authors, Year, Journal, Vol#, Page#
% Journal Title Abbreviation >> http://www.asj.or.jp/pasj/Jabb.html

\end{thebibliography}
\end{document}